\documentclass[twocolumn]{autart}

\usepackage{epsfig}

\usepackage{amssymb}

\usepackage{graphicx}

\usepackage{amsmath}

\usepackage{amssymb}

\usepackage{dcolumn}

\usepackage{bm}

\begin{document}

\begin{frontmatter}

\title{Oscillations of factorial cumulants to factorial moments ratio from an eikonal approach}

\author[urgs,uen]{P. C. Beggio}\ead{beggio@uenf.br}

\address[urgs]{Universidade Federal do Rio Grande do Sul - UFRGS,\\
  Porto Alegre - RS, Brazil.}

\address[uen]{Universidade
Estadual do Norte Fluminense Darcy Ribeiro - UENF,\\
Campos dos Goytacazes - RJ, Brazil.}


\begin{abstract}
We study the factorial moments ($F_{q}$), the factorial cumulants
($K_{q}$) and the ratio of $K_{q}$ to $F_{q}$ ($H_{q}=K_{q}/F_{q}$)
in $pp$/$p\overline{p}$ collisions using an updated approach, in which the
multiplicity distribution is related to the
eikonal function. The QCD inspired eikonal model adopted contains contributions of quark-quark, quark-gluon and
gluon-gluon interactions. Our work shows that the approach can
reproduce the collision energy dependence of the $F_{q}$ moments, correctly predicts that
the first minimum of the $H_{q}$ lies around $q$ $=$ 5 and qualitatively reproduces the oscillations of the $H_{q}$ moments, as shown in the
experimental data and predicted by QCD at preasymptotic energy. The result of this study seems to indicate that
the $H_{q}$ oscillations are manifestation of semihard component in the multiparticle production process.
Predictions for multiplicity distribution and $H_{q}$ moments at the LHC energy of $\sqrt{s}$ = 14 TeV are presented.

\end{abstract}

\end{frontmatter}

\section{Introduction}

The hadronization mechanism is one of the main problems of QCD \cite{Braz_Japn} and
the probability $P_{n}$ for producing $n$ charged particles in
final states (namely multiplicity distribution) is related to the
production mechanism of the particles.  It also contains
information about multiparticle correlations in an integrated form \cite{GUNPB} \cite{GUimpA} \cite{DreminGary}.
In the multiparticle production dynamics sector of strong
interactions, the multiplicity distribution $P_{n}$, its
factorial moments $F_{q}$ and its cumulant moments $K_{q}$ are
relevant physical observables \cite{GUNPB} \cite{GUimpA}. As pointed out in \cite{DreminAtomic},
moment analysis
of $P_{n}$ can also be performed for models as well as for experimental
data of hadronic processes. Hence, this approach is common to all
processes and methods of analysis.
The ratio of factorial cumulants to factorial moments, ($H_{q}=K_{q}/F_{q}$), turns out to be a
relevant physical object because, in the framework
of perturbative QCD, it was theoretically predicted that the $K_{q}$
(or equivalently the $H_{q}$ moments) of $P_{n}$ should present
oscillations in sign as a function of its order $q$, and that $H_{q}$
has its first minimum at $q$ $\approx$ 5
\cite{DreminNechitaiho}. The predictions turned out to be confirmed
by experiments in all known high energy processes ($e^{+}e^{-}$,
$pp$, $p\overline{p}$, hadron-nucleus and nucleus-nucleus
collisions) \cite{DiasdeDeus}. QCD yields results on partons, and not
hadrons, and since the partons interactions determine the
outcome of collisions with many particles produced, the moments
$H_{q}$ are interesting to study because calculations done at parton
($p$) level can be directly applied at hadron ($h$) level in the
framework of the Generalized Local Parton Hadron Duality (GLPHD), in
which we have $H_{q}^{(h)}$=$H_{q}^{(p)}$ (for details see
\cite{GUimpA}). However, the QCD has not yet been able to explain
satisfactorily the $P_{n}$ and the formation of hadron from their
constituent quarks and gluons. Thus, phenomenological approaches are
used as source of information for adequate theoretical developments. In \cite{BeggioNPA}, we have update and applied an existing
phenomenological approach, referred to as
\emph{Simple One String Model} - $SOSM$ \cite{BeggioMV}, which allows simultaneous
description of several experimental data from elastic and inelastic channels. This simple model is also able to predict $P_{n}$ from ISR to Collider
energies (30.4 - 900 GeV) without free parameters \cite{BeggioHama} \cite{BeggioBJP}. In addition, the model allows us to study the inelasticity in $pp/p\overline{p}$ reactions, for which experimental information is scarce \cite{BeggioNPA}. Our goals are to study the
collision energy dependence of $F_{q}$ moments and investigate whether is possible
predict, within the framework of the $SOSM$, the position of the first minimum of the $H_{q}$ moments and its subsequent oscillations at
higher orders, as qualitatively predicted by QCD at preasymptotic energies and, in addition, compare it to the measured data. The plan of the paper is as
follows. In Section 2 we briefly discuss
the main points of the $SOSM$, as well as their inputs. In the section 3 we
present the basic definitions of the
multiplicity distribution moments, applying the theoretical framework of the model and computing some moments
of the $P_{n}$. Results and discussions are in Section
4. The concluding remarks are the content of Section 5.

\section{A model for $P_{n}$}

The main Eqs. of the $SOSM$ are
\begin{equation}
 <n(s)>P_{n}(s)={\frac
  {\int d^2b\,{\frac{[1-e^{-2\,\chi_{I}(s,b)}]}{f(s,b)}}\,
    \psi^{(1)}\!\left(\frac{z}{f(s,b)}\right)}
  {\int d^2b\,[1-e^{-2\,\chi_{I}(s,b)}]}},
 \label{Phi}
\end{equation}
where
\begin{equation}
f(s,b)=\xi(s)\,[\chi_{I}(s,b)]^{2A}\,,
\end{equation}

\begin{equation}
 \xi(s)=\frac
  {\int d^2 b\,[1-e^{-2\,\chi_{I}(s,b)}]}
  {\int d^2 b\,[1-e^{-2\,\chi_{I}(s,b)}]\,[\chi_{I}(s,b)]^{2A}}\,,\
 \label{xi}
\end{equation}
and
\begin{equation}
 \psi^{(1)}(z)=2\,\frac{k^k}{\Gamma(k)}z^{k-1}e^{-kz}\,.
 \label{psi1}
\end{equation}
$\psi^{(1)}$ corresponds to the KNO form of the negative binomial
distribution (NBD) (or gamma distribution) \cite{Fiete}, $\sqrt{s}$ is the center-of-mass energy,
$<n>$ is the average multiplicity, $z=n/<n(s)>$
represents the usual KNO scaling variable, $b$ is the impact
parameter, $\Gamma$ is the usual gamma function, $\chi_{I}$ is the
imaginary part of the complex eikonal function and $ \xi(s)$ is
determined by usual normalization condition on $P_{n}$
\cite{BeggioMV}. The main points of the model are summarized as
follows:
\begin{enumerate}
\item $P_{n}(s)$ has been constructed by summing contributions (multiparticle creation) coming
from hadron-hadron collisions taking place at fixed value of $b$;

\item at each value of $b$, the multiplicity distribution of the secondaries, $\psi^{(1)}$, is narrow and follows a gamma distribution, which is characterized by the $k$
parameter. The fact that the observed $P_{n}$ is broad arise from
integration over $b$. There are both empirical and theoretical reasons for the choice of Eq. (4). The form successfully describes the empirical multiplicity distributions in $e^{+}$$e^{-}$ annihilations, yielding the values of $A$ and $k$ parameters adopted in this analysis \cite{BeggioMV}. Theoretically, the gamma distribution arise as a solution of the QCD evolution equation for three gluon branching \cite{Ina}, which is a good approximation, since at high energies gluons dominate the parton-parton cross section. More
recently NBD has been derived from the Color Glass Condensate approach, which indeed predict the needed NBD \cite{Gelis};

\item the energy deposited for particle production, at each $b$, depends upon the eikonal function in a power
law form,  \cite{BeggioNPA} \cite{BeggioMV} \cite{BeggioHama}
\cite{BeggioBJP}
\end{enumerate}

\begin{equation}
<n(s,b)>\,\propto\,[\chi_{I}(s,b)]^{2A}.
\end{equation}

The last Eq. is physically motivated by the fact that the
eikonal may be interpreted as an overlap, on the $b$ plane, of two
colliding matter distributions \cite{Barshay1982}. A possible physical picture of $SOSM$ is that a process,
occurring at fixed value of $b$, may be interpreted as due to an
elementary parton-parton collision, with the formation of an object like
a string \cite{BeggioHama} \cite{BeggioBJP}. Hence, probably one parton-parton pair
interaction has triggered the multitude of final particles observed
in the interval 30 $<$ $\sqrt{s}$ $\leq$ 900 GeV. The model permits a calculation of the full phase space $P_{n}$, as
well as its moments, once the eikonal is given and appropriate values
for $k$ and $A$ are adopted. In this work we have kept the choices
done in \cite{BeggioMV}, borrowing their results, also used in
\cite{BeggioNPA}. Specifically, we have used the values of
$k$=10.775 and $A$=0.258 and adopted the QCD inspired complex
eikonal determined from elastic channel studies \cite{Block}, in
which the eikonal is written as a combination of an even and odd
eikonal terms related by crossing symmetry
$\chi_{pp}^{\overline{p}p}{(s,b)}=\chi^{+}{(s,b)}\pm\chi^{-}{(s,b)}$.
The even eikonal is written as the sum of quark-quark, quark-gluon
and gluon-gluon contributions:
\begin{equation}
\chi^{+}{(s,b)}=\chi_{qq}{(s,b)}+\chi_{qg}{(s,b)}+\chi_{gg}{(s,b)}\,,
 \label{sigma}
\end{equation}
while the odd eikonal is parametrized as
\begin{equation}
\chi^{-}{(s,b)}=C^{-}\sum\frac{m_{g}}{\sqrt{s}}e^{i\pi/4}W(b;\mu^{-}).
 \label{sigma1}
\end{equation}
The various parameters and functions involved in last two equations
are discussed in \cite{Block}, and the adopted QCD-inspired eikonal satisfies crossing symmetry, analyticity and unitarity
\cite{Block}.

\section{Analysis of experimental data}

The normalized factorial moments
are defined by \cite{Fiete}
\begin{equation}
F_{q}=\frac{1}{<n>^{q}} \sum_{n=q}^{\infty}
n(n-1)...(n-q+1)\,P_{n}\,, \label{fatorial}
\end{equation}
where $<n>=\sum nP_{n}$ is the average multiplicity and $q$ is a
positive integer and represent the order (or rank) of the moments.
The normalized factorial cumulants can be calculated
recursively from $F_{q}$ according to
\begin{equation}
F_{q}=\sum_{m=0}^{q-1} \frac{(q-1)!}{m!(q-m-1)!}\,K_{q-m}\,F_{m}\,.
 \label{cumulant}
\end{equation}

In turn, the ratio of factorial cumulants to factorial moments of $P_{n}$, are
written as
\begin{equation}
H_{q}=\frac{K_{q}}{F_{q}}\,
\label{hq}
\end{equation}
and should present oscillations in sign as a function of its order $q$. Some experimental results, used in this work, are given in terms of the reduced $C$-moments and defined by \cite{Fiete}
\begin{equation}
C_{q}=\frac{<n^{q}>}{<n>^{q}}=\frac{\sum n^{q}P_{n}}{(\sum nP_{n})^{q}}\,,
 \label{sigma1}
\end{equation}
where $q$ is a positive integer. The normalized $C$-moments can be
expressed through normalized factorial moments $F_{q}$ and, for the
first five moments we have \cite{Michal}
\begin{equation}
C_{2}=\frac{1}{<n>}+F_{2},
\label{c2}
\end{equation}
\begin{equation}
C_{3}=\frac{3\,C_{2}}{<n>}-\frac{2}{<n>^{2}}+F_{3}, \label{c3}
\end{equation}
\begin{equation}
C_{4}=\frac{6\,C_{3}}{<n>}-\frac{11\,
C_{2}}{<n>^{2}}+\frac{6}{<n>^{3}}+F_{4}, \label{c4}
\end{equation}
\begin{equation}
C_{5}=\frac{10\,C_{4}}{<n>}-\frac{35\,C_{3}}{<n>^{2}}+\frac{50\,
C_{2}}{<n>^{3}}-\frac{24}{<n>^{4}}+F_{5}. \label{c5}
\end{equation}

The $F_{q}$ and $H_{q}$ moments of charged particles at the CERN ISR in $pp$
collisions and of the UA5 Collaboration in $p\overline{p}$ are
analyzed by $SOSM$. The Eqs. (1) - (4) have been applied to obtain
the theoretical values of $P_{n}$. The $F_{q}$, $K_{q}$  and $H_{q}$
moments were obtained from theoretical and experimental values of
$P_{n}$ according to Eqs. (8) - (10), where the known values of
$F_{1}$=$K_{1}$=1 and $F_{0}$=0 have been used \cite{DreminGary}. As
mentioned, the model enable us to calculate $P_{n}$ in full
phase space. Thus, at ISR energies (30.4 - 62.2 GeV) and at Collider energy of 546 GeV, we have studied the sample of inelastic events \cite{ABC} \cite{UA51}, and studied the non-single-diffractive (NSD) experimental data at energies of 200 and 900 GeV since, for the last two energies, $P_{n}$ have been measured
only in NSD $p\overline{p}$ collisions. \cite{CERN200_900}.
The $\sqrt{s}$ dependence of the $F_{q}$ moments, from both model and
experimental data, are compared in Fig. 1. Prediction for $F_{q}$ at 14 TeV is also presented in Fig. 1. We now turn to
calculation of the $H_{q}$ moments as a function of its order $q$. The
results are shown at ISR energies in Figs. 2-5 and those for the UA5
Collaboration, from 200 to 900 GeV, in Figs. 6-8. For the sake of
discussions, we have also plotted the corresponding $P_{n}$ against
$n$. Theoretical $P_{n}$ and $H_{q}$ plots in full phase space at the LHC energy of $\sqrt{s}$=14 TeV are presented in Fig. 9, in
which we have used the value of $<n(s)>$ $\simeq$ 108 from Ref. \cite{Troshin}. Here,
we would like to observe that, at LHC, $P_{n}$ in $pp$ collisions at $\sqrt{s}$ = 0.9, 2.36 and 7 TeV have been measured in different pseudorapidity ranges,
from $|\eta| <$ 0.5 to $|\eta| <$ 2.4 \cite{CMS}. However, the $SOSM$ enable us to calculate $P_{n}$, and its moments, in full phase
space, $|\eta| <$ 5. For this reason, comparison of model predictions to LHC data are not presented.

\section{Results and discussions}

The energy dependence of the $F_{q}$ can be used to study the KNO
scaling hypothesis \cite{Fiete}. The model predicts that for each value of $q$, $F_{q}$ increases only slightly over the
ISR range, but increases more substantially between the ISR and the
CERN Collider, in good agreement with the experiments, indicating that the KNO scaling is
approximately valid at ISR, Fig. 1. However, as the energy rises into the CERN
Collider and LHC ranges, $F_{q}$ shows an increase with $\sqrt{s}$,
demonstrating that the KNO scaling is broken. $P_{n}$ in the Eq.
(1) depends on the eikonal and also on the values of $A$ and $k$
parameters, which are fixed numbers ($A$=0.258, $k$=10.775) in the interval of energy studied. Hence, the model relates the energy
dependence of the $F_{q}$ to that of the eikonal function. Thus, the $SOSM$ qualitatively predicts the
violation of KNO scaling by changing only the eikonal function, without changing the underlying
elementary interaction, in agreement with what could be expected from QCD.
This result is qualitatively in agreement with those obtained by CMS Collaboration, albeit in smaller pseudorapidity intervals
\cite{CMS} \cite{BeggioNPA}.
It seems interesting, since the adopted QCD inspired eikonal, Eqs. (6) - (7), has been determined phenomenologically to give good fits
to measurements of elastic channel observables, namely the total cross section
($\sigma_{tot}$), the ratio of the real to the imaginary part of the
forward scattering amplitude ($\rho$), and the logarithmic slope of
the differential elastic scattering cross section in the forward
direction ($B$) \cite{Block}. Concerning the $H_{q}$ moments, at the energies of $\sqrt{s}$ = 52.6, 546 and 900 GeV model
predictions are in good agreement with the data up to $q$ = 16. Now, at the energies of $\sqrt{s}$ = 30.4,
44.5 and 62.2 GeV the data behavior seems theoretically described
until $q$ $\approx$ 13. At $\sqrt{s}$ = 200 GeV there are discrepancies between theoretical and observed points. However, we can see that the
theoretical first minimum lies around $q$=5, for all energies studied in this work. At this point we would like to call attention for
the interesting results: the $SOSM$ qualitatively reproduces the $H_{q}$
oscillations and correctly predicts the position of the first minimum, in wide interval of energy
and with no free parameter, which could be adjusted to achieve agreement with the experimental results. Now, we proceed to a possible
interpretation of the $H_{q}$ oscillations in the framework of $SOSM$. However, it should be noted
that the oscillatory observed behavior of $H_{q}$'s has one
unphysical origin, which is consequence of dealing with truncated
$P_{n}$, due to finite statistics of data samples
\cite{Ugo_Gio_Lupia}. In the present approach, the even part of the eikonal, Eq. (6), contains
quark-quark, quark-gluon and gluon-gluon components, being the
eikonal functions $\chi_{qq}(s,b)$ and $\chi_{qg}(s,b)$ needed to
describe the lower energy forward data, while $\chi_{gg}(s,b)$
contribution dominates at high energy and determines the asymptotic
behavior of cross sections. The rise of the cross section with $\sqrt{s}$ is
consequence of the increasing number of gluons soft populating the colliding particles, increasing
the probability of perturbative small-$x$ gluon-gluon collisions. The multiparton interactions lead to the appearance of
minijets, implying in the semihard contribution in the multiparticle production process, as $\sqrt{s}$ increases.
It is also important to note that the eikonal $\chi_{gg}(s,b)$, used by Block et al., is identical to the one used in mini-jet models
\cite{Block}. Hence, and in view of Eq. (5), the physical picture that emerges is that at relatively lower collision energy,
the particle production process is characterized by soft events, receiving
contributions from $qq$, $qg$ and $gg$ scattering events. At high energy, where the $\chi_{gg}(s,b)$ component is predominant, the
particle production process is dominated by semihard scattering events, as
can be inferred from the increase in the number of gluons populating the incident particles, with the appearance of minijets.
Hence, since the general tendency of quite complicated oscillatory behavior of the $H_{q}$ moments are
qualitatively well reproduced by the $SOSM$, it seems quite natural to ascribe it as being manifestations of the semihard
component in the multiparticle production mechanism.

\section{Concluding Remarks}

By using an eikonal approach which allows simultaneous description of several
experimental data from elastic and inelastic channels, through unitarity equation in the
impact parameter representation, we have analyzed the dependence of the $F_{q}$ moments as a function of $\sqrt{s}$, the
oscillatory behavior of the $H_{q}$ moments and the position of its first minimum. Through the specific form of the multiplicity distribution for the secondaries, Eq. (4),
a connection is made between the geometric phenomenological approach and the underlying theory of parton branching.
It is important to note that the same approach was used to investigate both, the impact
parameter and collision energy dependence of the inelasticity in $pp/p\overline{p}$ reactions \cite{BeggioNPA}.
Thus, we would like to emphasize the main results from our studies: The approach reproduces the
collision energy dependence of the $F_{q}$ moments, which comes from elastic channel
analysis, satisfactorily reproduces the behavior of the $H_{q}$ moments and correctly predicts the position of the first minimum, in agreement with the predictions from higher order perturbative QCD.
From the increase in the content of gluons into protons, leading to appearance of minijets, we have inferred that the oscillatory
behavior of the $H_{q}$ moments are manifestations of the semihard component in the multiparticle production process. We have
shown the consistency between model and
QCD results which, from our understanding, is an important step to encourages future improvements in
the approach that we have discussed. Thus, in this sense, the present work complements the previous analysis done in \cite{BeggioNPA}, since
moment analysis of $P_{n}$ is common to all hadronic processes and methods of analysis \cite{DreminAtomic}.

\begin{ack}
I am grateful to E.G.S. Luna for discussions.
\end{ack}


\begin{figure}[hb!]
\begin{center}
\includegraphics*[height=9.5cm,width=8.4cm]{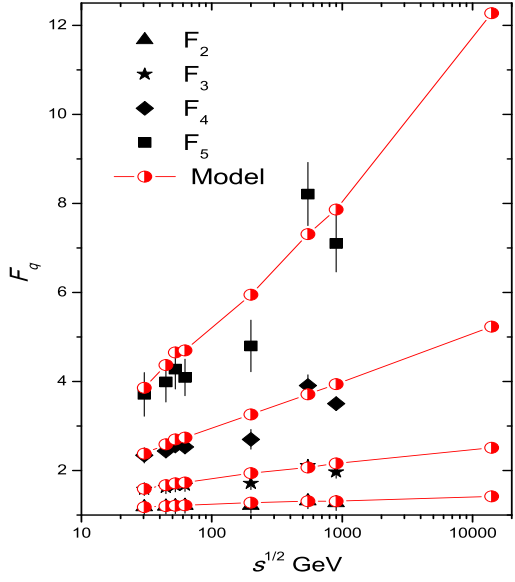}
\caption{Normalized factorial moments as a function of the collision
energy. At energies of $\sqrt{s}$ = 200 and 900 GeV the data points are from NSD events
\cite{CERN200_900}. At other energies the data are from inelastic
events \cite{ABC} \cite{UA51}. Prediction for $F_{q}$ at $\sqrt{s}$= 14 TeV is presented. The lines are draw only as a guidance
of the theoretical points.} \label{fused445}
\end{center}
\end{figure}

\begin{figure}[ht!]
\begin{center}
\includegraphics*[height=9.5cm,width=8.4cm]{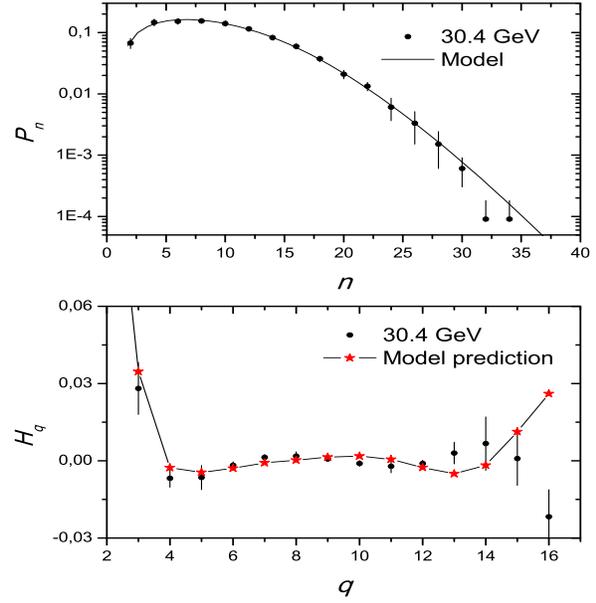}
\caption{Theoretical and experimental plots of $P_{n}$ against $n$ and $H_{q}$ as a function of its order $q$
for $pp$ at ISR energy. Data points are from \cite{ABC} and the
lines are draw only as a guidance (n=2 to 34).} \label{fused445}
\end{center}
\end{figure}

\begin{figure}[h!]
\begin{center}
\includegraphics*[height=9.5cm,width=8.4cm]{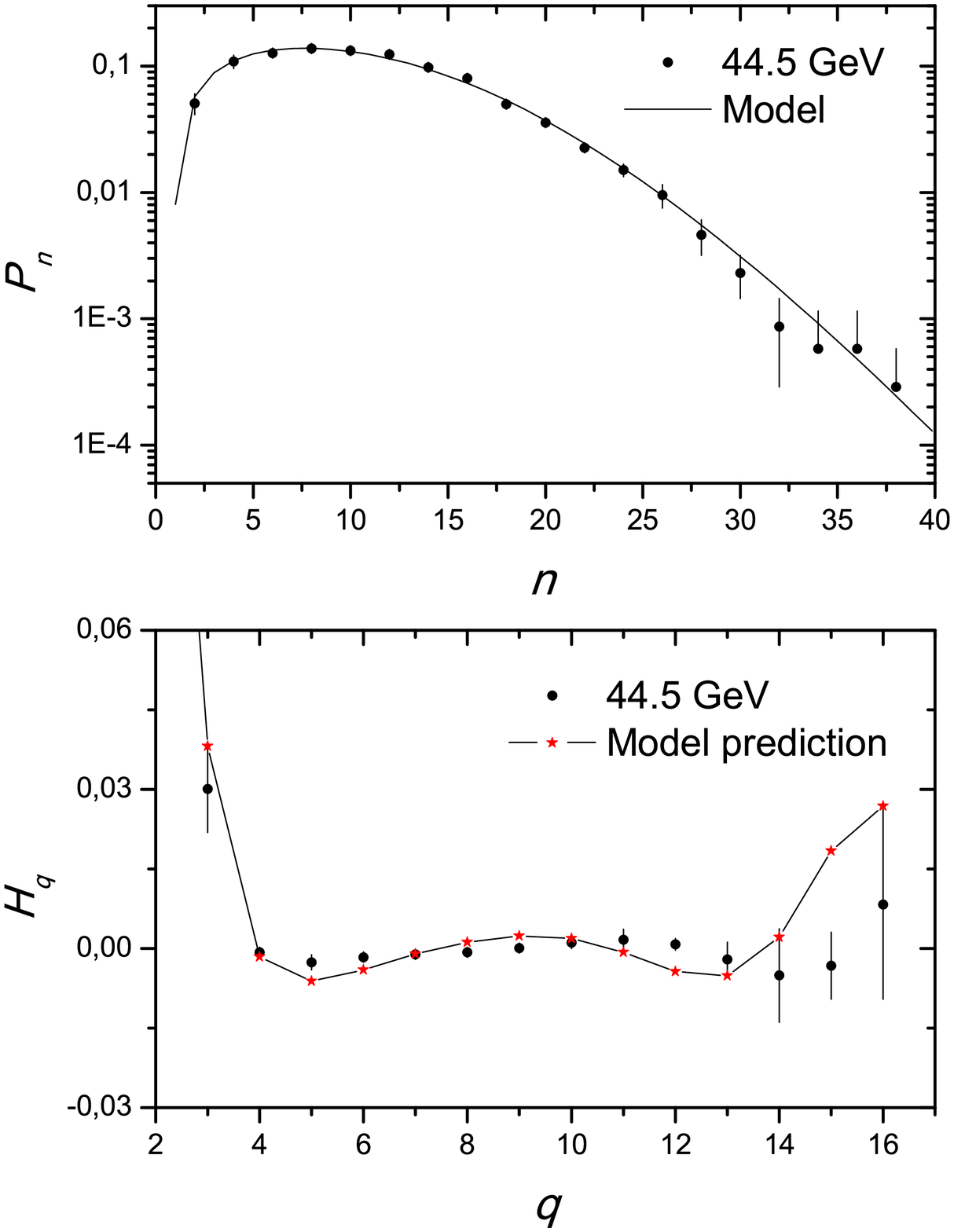}
\caption{Theoretical and experimental plots of $P_{n}$ against $n$ and $H_{q}$ against $q$
for $pp$ at ISR energy. Data points are from \cite{ABC} and the
lines are draw only as a guidance (n=2 to 38).} \label{fused445}
\end{center}
\end{figure}

\begin{figure}[h!]
\begin{center}
\includegraphics*[height=9.5cm,width=8.4cm]{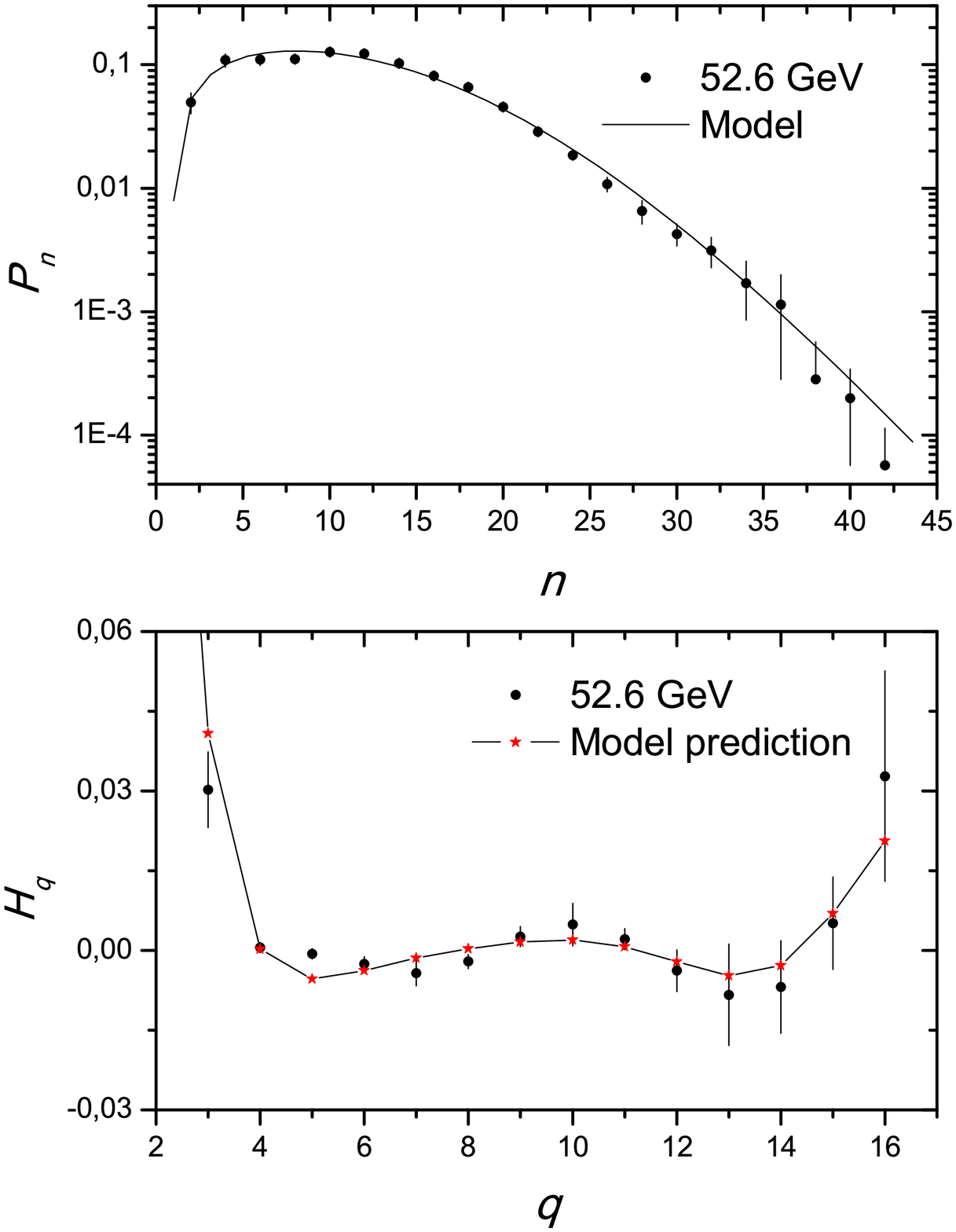}
\caption{Theoretical and experimental plots of $P_{n}$ against $n$ and $H_{q}$ against $q$
for $pp$ at ISR energy. Data points are from \cite{ABC} and the
lines are draw only as a guidance (n=2 to 42).}
 \label{fused445}
\end{center}
\end{figure}

\begin{figure}[h!]
\begin{center}
\includegraphics*[height=9.5cm,width=8.4cm]{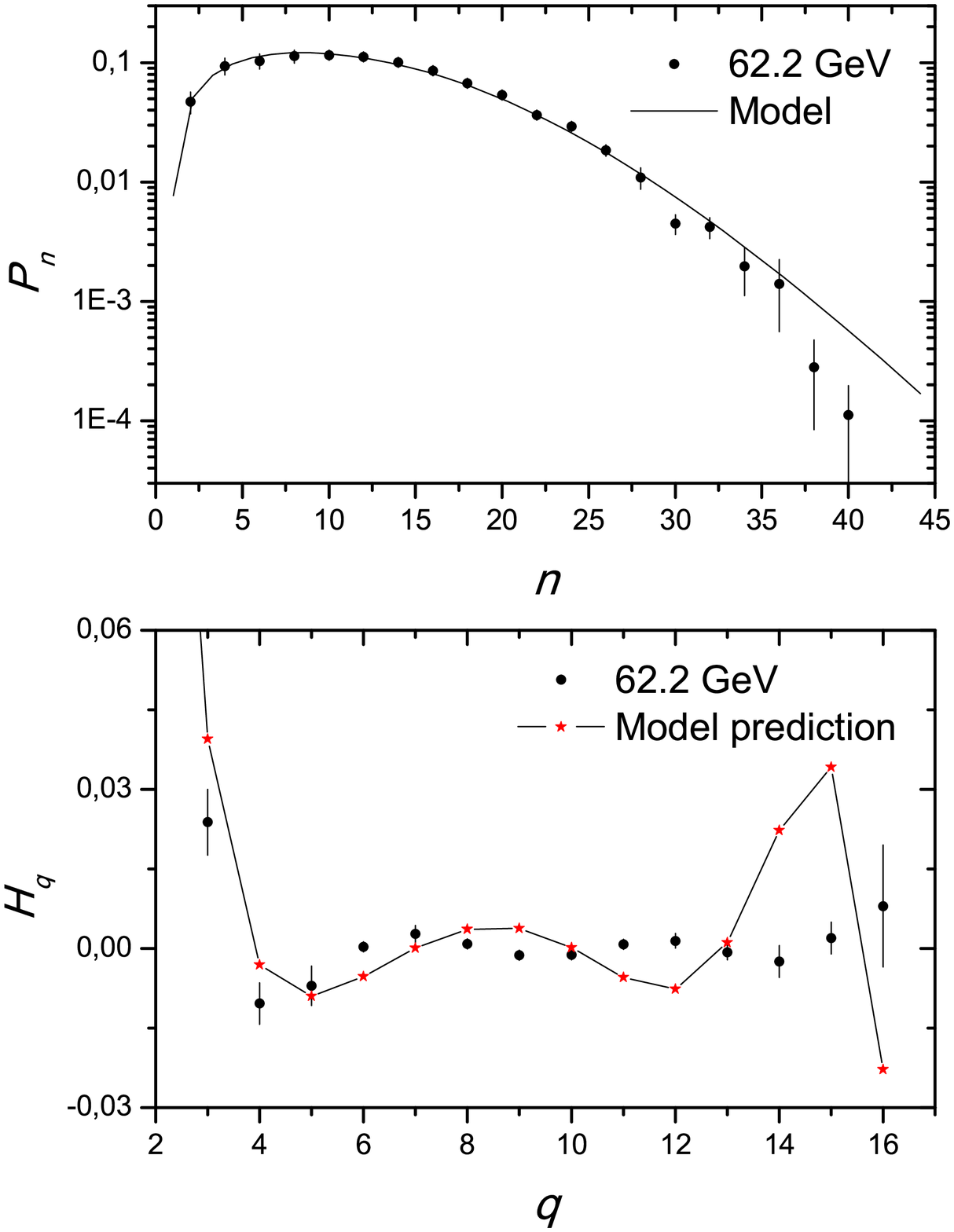}
\caption{Theoretical and experimental plots of $P_{n}$ against $n$ and $H_{q}$ against $q$
for $pp$ at ISR energy. Data points are from \cite{ABC} and the
lines are draw only as a guidance (n=2 to 40).}
 \label{fused445}
\end{center}
\end{figure}

\begin{figure}[htt!]
\begin{center}
\includegraphics*[height=9.5cm,width=8.4cm]{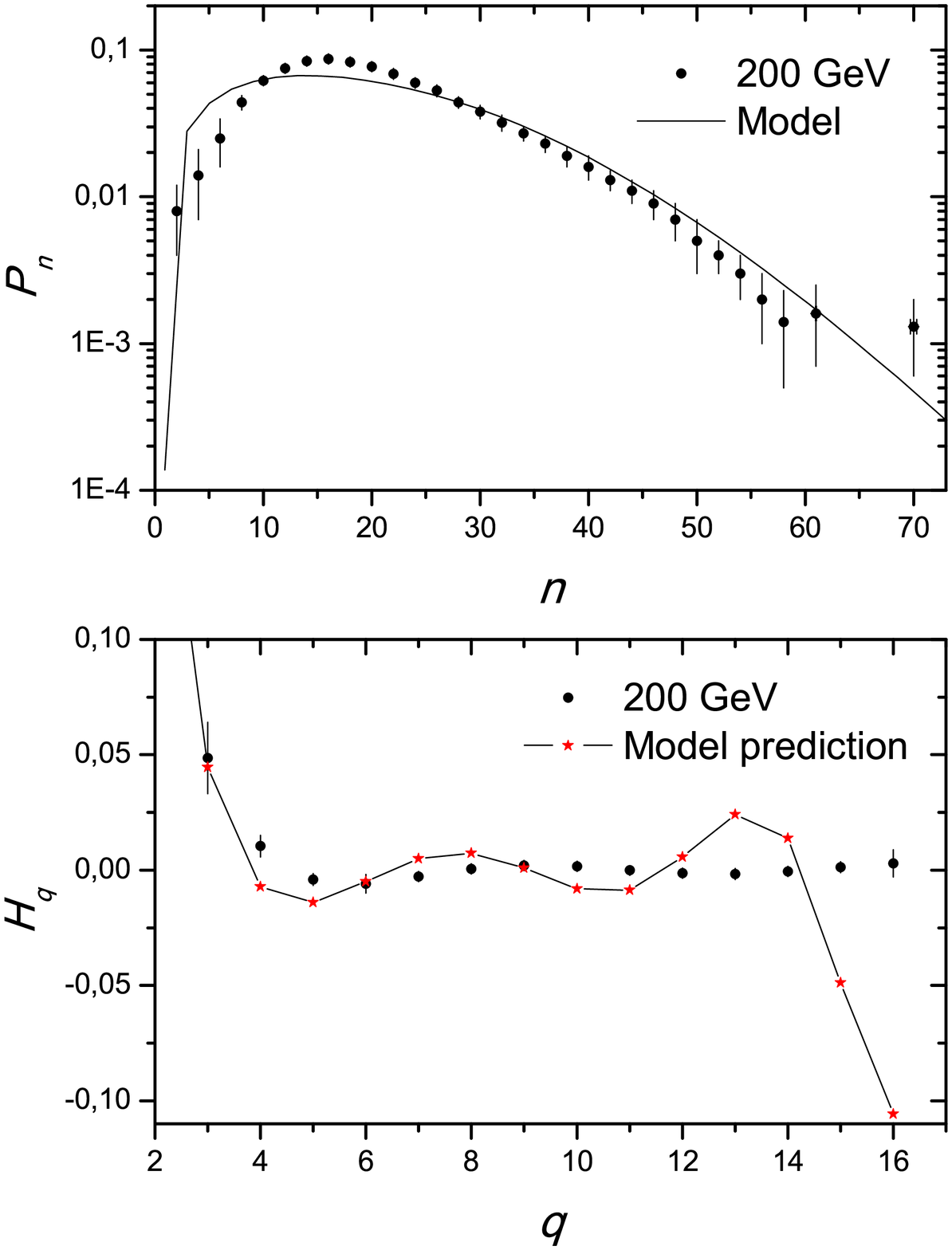}
\caption{Theoretical and experimental plots of $P_{n}$ against $n$ and $H_{q}$ against $q$
for $p\overline{p}$ at Collider energy. Data points are from
\cite{CERN200_900} and the lines are draw only as a guidance (n=2 to
70).} \label{fused445}
\end{center}
\end{figure}

\begin{figure}[htt!]
\begin{center}
\includegraphics*[height=9.5cm,width=8.4cm]{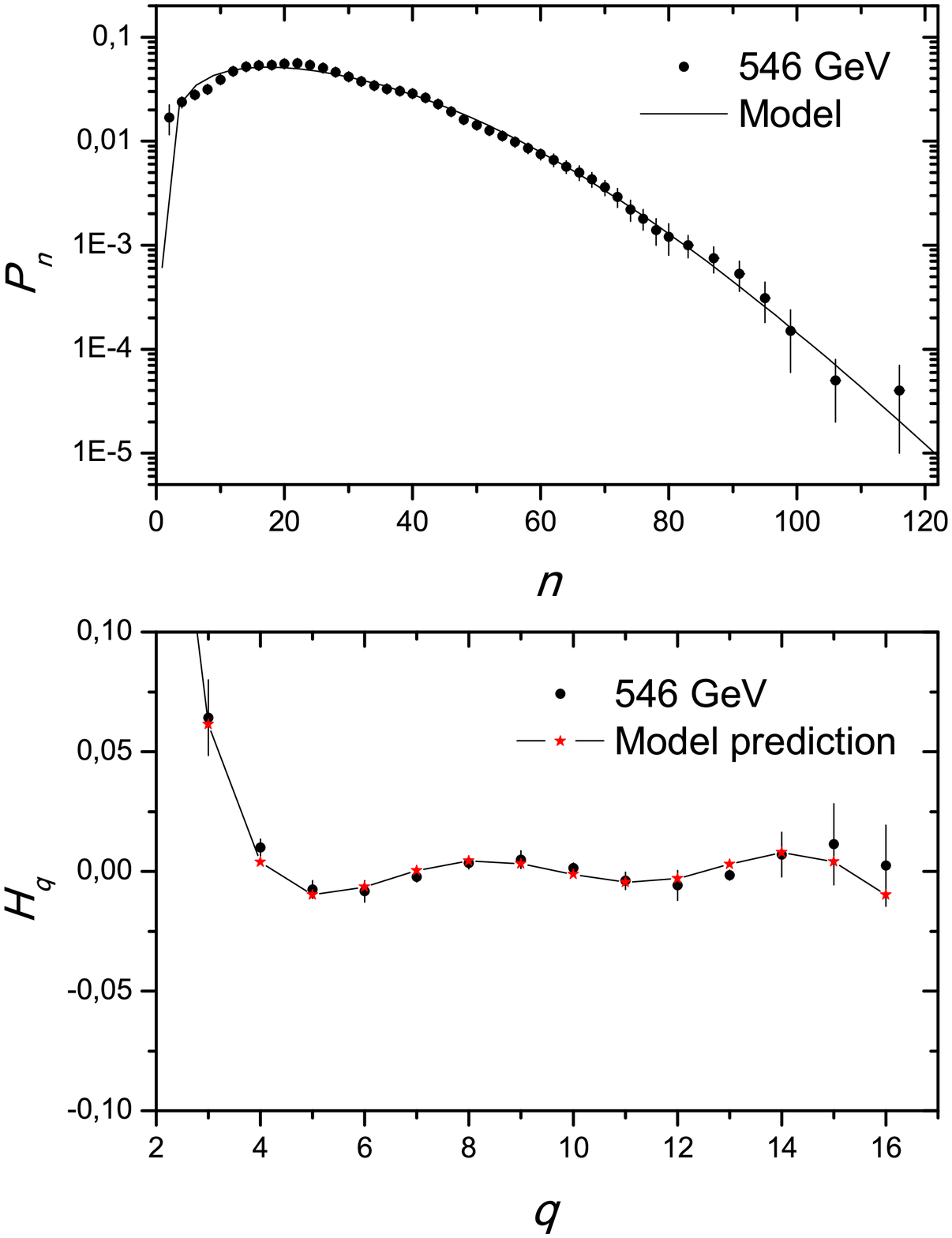}
\caption{Theoretical and experimental plots of $P_{n}$ against $n$ and $H_{q}$ against $q$
for $p\overline{p}$ at Collider energy. Data points are from
\cite{UA51} and the lines are draw only as a guidance (n=2 to 116).}
\label{fused445}
\end{center}
\end{figure}

\begin{figure}[h!]
\begin{center}
\includegraphics*[height=9.5cm,width=8.4cm]{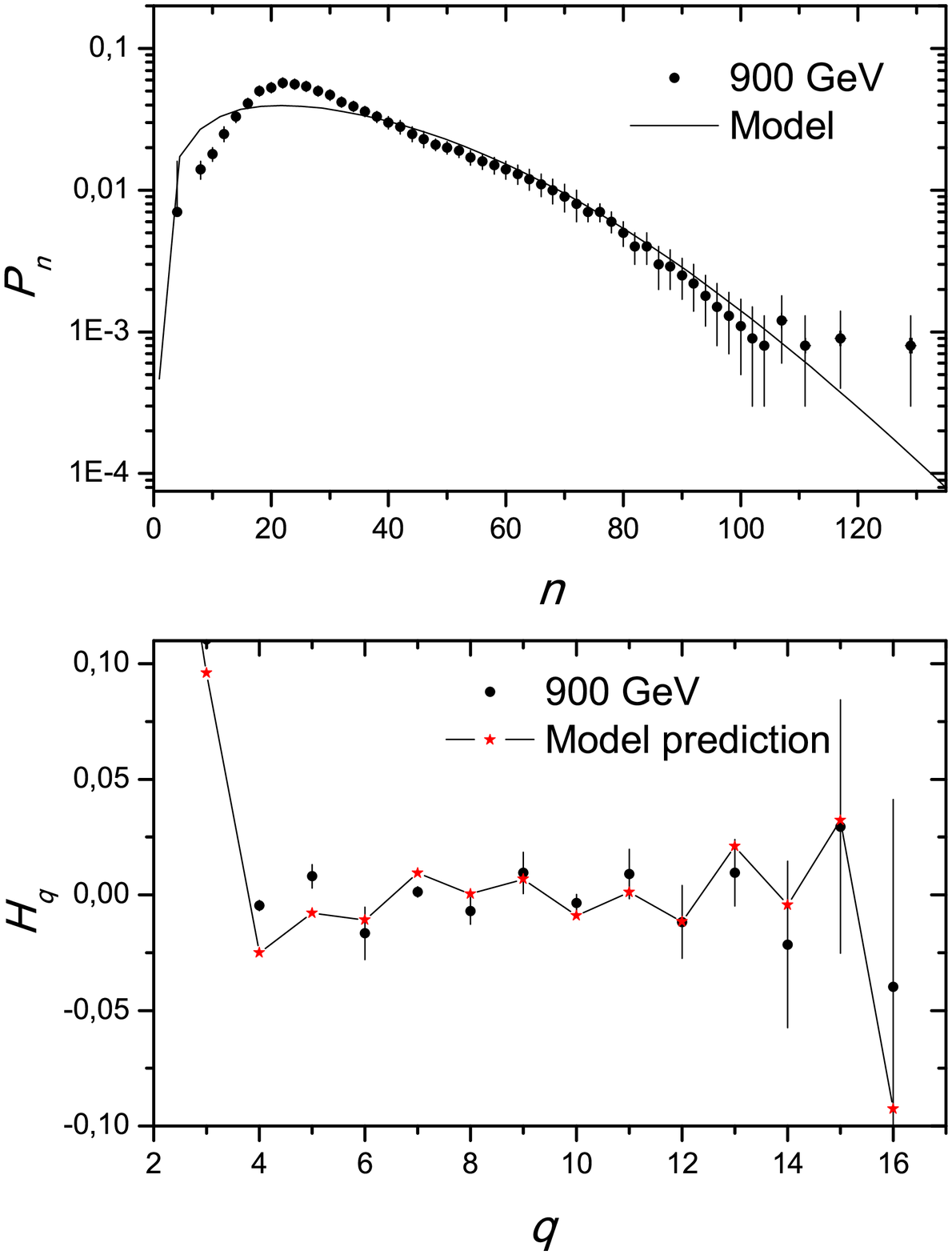}
\caption{Theoretical and experimental plots of $P_{n}$ against $n$ and $H_{q}$ against for
$p\overline{p}$ at Collider energy. Data points are from
\cite{CERN200_900} and the lines are draw only as a guidance (n=4 to
129).} \label{fused445}
\end{center}
\end{figure}

\begin{figure}[h!]
\begin{center}
\includegraphics*[height=9.5cm,width=8.4cm]{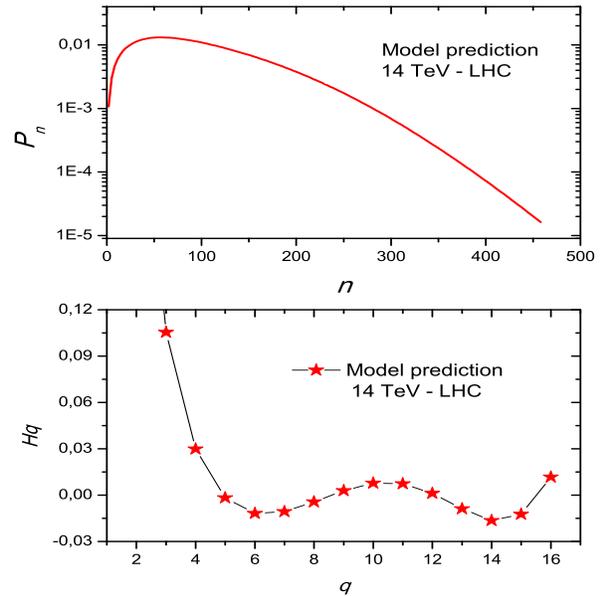}
\caption{Theoretical plots of $P_{n}$ against $n$ and $H_{q}$ against $q$ for $pp$ at LHC
energy. The line is draw only as a guidance (n=4 to 350).}
\label{fused445}
\end{center}
\end{figure}

\end{document}